\documentclass[prl,twocolumn,superscriptaddress]{revtex4-2}   
\usepackage{graphicx,amssymb}
\usepackage{color,ulem}
\usepackage{bm}   

\usepackage{bm}% bold math
\usepackage{amssymb,amsmath}
\usepackage{amsthm}

\usepackage{latexsym}
\usepackage{graphicx,amssymb}
\usepackage{color,ulem}
\usepackage{float}
\usepackage{siunitx}
\usepackage{mathrsfs} 
\usepackage{soul}
\usepackage{graphicx}% Include figure files
\usepackage{dcolumn}% Align table columns on decimal point
\usepackage{bm}% bold math
\usepackage[utf8]{inputenc}
\usepackage{amsfonts,amssymb,amsmath}
\usepackage[mathscr]{eucal}
\usepackage{epsfig}

\begin{document}
% \preprint{AIP/123-QED}

\title{Thermalization via three-wave mixing}%\\Bolometric ac to dc conversion of Josephson current\\Bolometric detection of ac Josephson current\\Bolometric detection of Josephson current - ac to dc conversion\\Bolometric rectification of (ac) Josephson current\\Rectification of ac Josephson current by a bolometer
% \thanks{A footnote to the article title}
\author{Jukka P. Pekola}
\author{Bayan Karimi}
\affiliation{Pico group, QTF Centre of Excellence, Department of Applied Physics, Aalto University, P.O. Box 15100, FI-00076 Aalto, Finland}

\date{\today}

\begin{abstract}
%We discuss thermalization in a multimode quantum cavity. An isolated cavity with quadratic couplings does not thermalize. We find that a generic three-wave perturbation, typical for instance in superconducting Josephson systems, leads to thermalization into a Bose-Einstein distribution of occupations of the modes. The temperature of this state is dictated by energy conservation in this closed system. The thermal state is robust against weak disturbances. We discuss possible physical realizations of the effect in solid-state systems.
We discuss thermalization in a multimode quantum cavity under unitary evolution. According to general principles, an isolated system with quadratic couplings does not exhibit thermalization. However, we find that three-wave perturbation, typical for instance in superconducting Josephson systems, may lead to thermalization into a Bose-Einstein distribution of occupations of the modes. The temperature of this state is dictated by energy conservation in this closed system, and the thermal distribution is robust against weak disturbances. We discuss how our findings open up new avenues to experimentally probe fundamental assumptions of statistical physics in solid-state systems.
 \end{abstract}
\maketitle
%The question of thermalization and temperature in isolated quantum systems has attracted lots of attention over the past years.
Thermalization has been a fundamental issue since the birth of statistical physics, and for quantum systems since early 20th century. 
Although dominated by theory \cite{abanin2019,mori2018,rigol2016,huse2015}, experiments especially on cold atoms have addressed this problem as well \cite{kinoshita2006,gring2012,kaufman2016}. Solid state provides likewise tailored quantum systems, e.g., in form of superconducting quantum circuits \cite{oliver2020}  and mesoscopic electronic structures in semiconductors, metals and two-dimensional materials \cite{heikkila2013}. Yet, up to now realizing sufficiently isolated solid-state systems for studies of internal thermalization has not been feasible in most of these systems. There are exceptions, however: for instance electrons in metals at low temperatures form a Fermi-Dirac distribution within a relaxation time of about 1 ns \cite{pothier1997} after a "quench", whereas these electrons relax to the external bath formed by lattice phonons over time-scales of about 100 $\mu$s \cite{RMP2006}. Such a huge separation of time-scales allows one to assume electrons to form an isolated interacting quantum system that reaches a quasi-equilibrium thermal state. Another class of solid-state systems becoming suitable for studies of quantum thermalization is formed of superconducting circuit QED (cQED) structures \cite{neill2016,chen2021,sundelin2024}. With the very rapid development of quantum information processing devices based on superconducting resonators and qubits, the degree of isolation measured by high quality factors of resonators and long coherence times of qubits now allows us to consider these systems also as quasi-isolated ones. This development opens a new exciting prospect for experimental studies of quantum thermalization.

\begin{figure}
	\centering
	\includegraphics [width=0.5\textwidth] {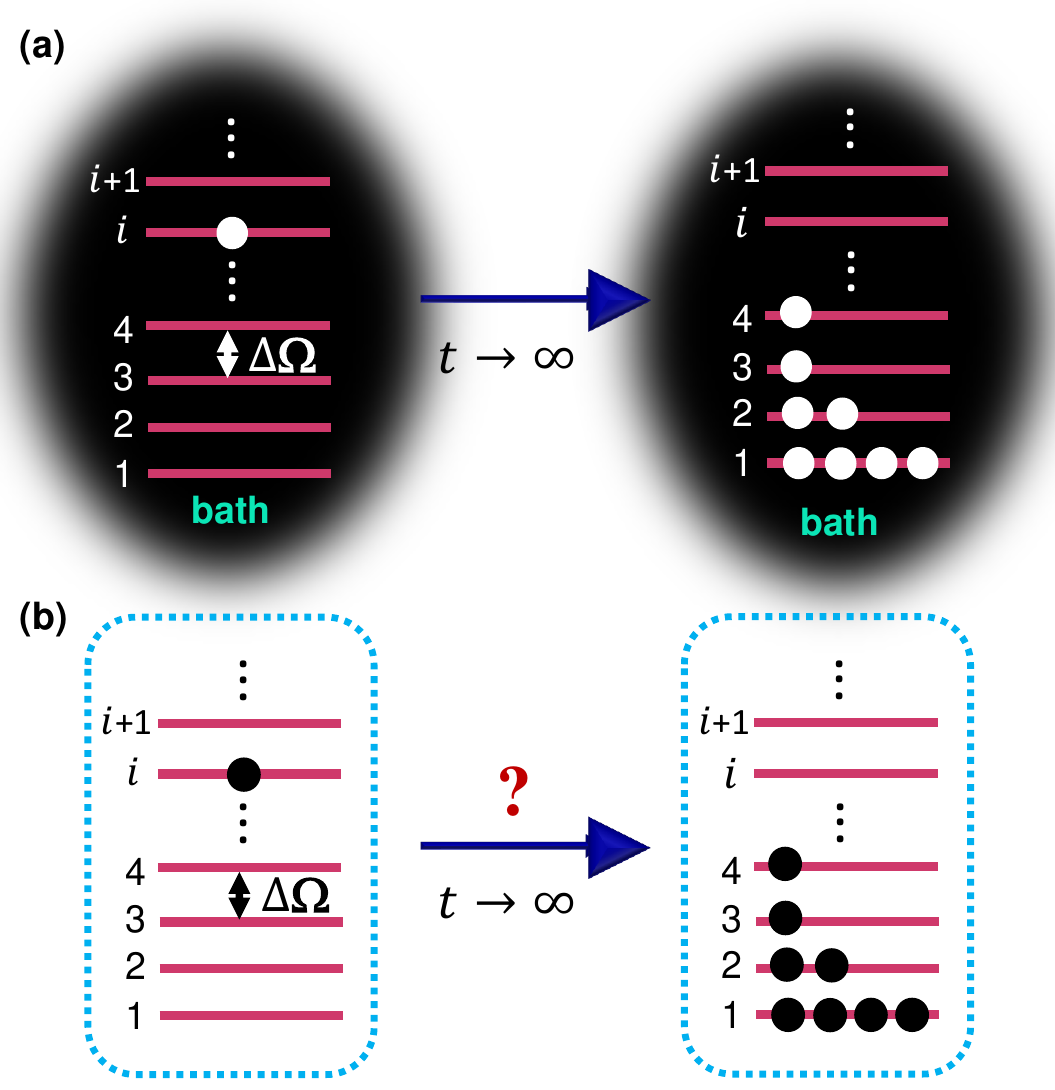}
	\caption{The dynamics of a multimode quantum resonator. (a) Resonator coupled to a heat bath reaches a thermal state in the long-time limit. (b) Here we study the population dynamics of an isolated resonator. 
		\label{fig1}}
\end{figure}

In this paper we demonstrate how a realistic nearly integrable system reaches a thermal state due to non-linearity in form of three-wave mixing. We consider a multimode cavity (see Fig. \ref{fig1}) that can be made of a superconducting coplanar wave resonator or alternatively of an array of Josephson junctions. Ideally its Hamiltonian is quadratic, which makes it integrable with the number of photons being a conserved quantity. Yet recent developments in cQED area have introduced elements that, combined with the linear cavities, lead to three-wave mixing type nonlinearities in the circuit. These nonlinearities can be caused by adding shifted cosine potentials due to asymmetries in superconducting Josephson junction loops in magnetic field, like in SNAIL (Superconducting Nonlinear Asymmetric Inductive eLement) configuration \cite{frattini2017}. Another recent strategy for three-wave mixing has been achieved by coupling a fluxonium qubit to a Josephson junction array \cite{manucharyan2023}. Naturally, a cavity coupled to a thermal bath as in Fig. \ref{fig1} (a) gets thermalized, but here we study explicitly an isolated cavity, Fig. \ref{fig1} (b).  

We focus on a generic multimode cavity whose Hamiltonian $\hat{\mathcal{H}}$ can be written as
\begin{equation}\label{Hamilton1}
	\hat{\mathcal{H}}=\hat{\mathcal{H}}_{0}+\hat V,
\end{equation}
where $\hat{\mathcal{H}}_{0}=\sum_{k=1}^N \hbar \Omega_k a_k^\dagger a_k$ is the non-interacting Hamiltonian. Here $N$ is the number of available modes, and $a_k^\dagger, a_k$ are the (bosonic) creation and annihilation operators of the $k$:th mode, respectively. The coupling between the modes, or for an open system that between a mode and environment, is described by $\hat V$. We take the eigen-energies of $\hat{\mathcal{H}}_{0}$ to be equidistant so that $\Omega_k = k \Delta\Omega$, where $\Delta\Omega$ is the free spectral range.

We are going to look at the evolution of the population operator for the $j$:th level, $\hat{N}_j=a_j^\dagger a_j$, and especially its expectation value $n(\Omega_j)\equiv \langle \hat{N}_j\rangle$. To find the kinetic equation for $n(\Omega_j)$, we write for the time rate of change of the operator $\hat{N}_j$, $\dot {\hat N}_j =\frac{i}{\hbar}[\hat V,\hat N_j]$, and use the Kubo formula in the interaction picture to obtain
\begin{equation}\label{kin1}
	\dot n(\Omega_j) =-\frac{i}{\hbar}\int_{-\infty}^0 dt'\langle [\dot {\hat N}_j,\hat V_I(t')]\rangle_0,
\end{equation}
where $\hat V_I(t) = e^{i\hat{\mathcal{H}}_{0}t/\hbar} \hat V e^{-i\hat{\mathcal{H}}_{0}t/\hbar}$, and $\langle \cdot\rangle_0$ denotes averaging in the non-interacting system.
 
Now the evolution of $n(\Omega_j)$ depends critically on the type of coupling $\hat V$ (besides the initial state of the cavity). If the modes are not coupled ($\hat V=0$), the populations $n(\Omega_j)$ are trivially stationary, and determined only by the initialization. Also, unsurprisingly, if the coupling is quadratic of the form
\begin{equation}\label{twowave}
	\hat V=\frac{1}{2}\sum_{p\neq q} g_{pq}(a_p+a_p^\dagger)(a_q+a_q^\dagger),
\end{equation} 
where $g_{pq}$ are arbitrary coupling constants between modes $p$ and $q$, there is no thermalization. This is expected from general arguments. We find this explicitly for our perturbative model as well, since the populations do not vary in time, i.e. $\dot n(\Omega_j)\equiv 0$, presented in the Supplemental Material. 

On the contrary, ubiquitous non-linearities do play a role in thermalization. In multi-wave mixing, energy conservation can be sequred in transitions unlike in coupling two modes only (Eq. \eqref{twowave}). Experimentally common examples are three- and four-wave mixing. In basic three-wave mixing, a photon is down-converted into two, or vice versa. Formally the coupling then reads 
\begin{equation}\label{threewave}
	\hat V=\sum_{p,q,r} (M_{pqr}a_p^\dagger a_q^\dagger a_r +M_{pqr}^* a_r^\dagger a_q a_p).
\end{equation}
In what follows, we write the couplings in the form $M_{pqr} \equiv A_{pq}B_r$ for convenience and for physical reasons. We then find the kinetic equation using Eq. \eqref{kin1} as
\begin{widetext} 
\begin{eqnarray}\label{kinfull}
	\dot n(\Omega_j) = &&\frac{4\pi}{\hbar^2}\nu_0\big\{\sum_{k=1}^N 2|A_{jk}|^2|B_{j+k}|^2[n(\Omega_j+\Omega_k)+n(\Omega_j)n(\Omega_j+\Omega_k)+n(\Omega_k)n(\Omega_j+\Omega_k)-n(\Omega_j)n(\Omega_k)]	\nonumber\\&& -\sum_{k=1}^{j-1} |A_{k,j-k}|^2|B_j|^2[n(\Omega_j)+n(\Omega_j)n(\Omega_k)+n(\Omega_j)n(\Omega_j-\Omega_k)-n(\Omega_k)n(\Omega_j-\Omega_k)]\big\}.
\end{eqnarray}
\end{widetext}
For derivation of Eq. \eqref{kinfull}, see Supplemental Material.
Here $\nu_0 =1/\Delta \Omega$ is the density of modes. We find easily that a Bose-Einstein distribution $n_B(\Omega)$ is a stationary solution of Eq. \eqref{kinfull}, i.e.
\begin{equation}\label{stationary}
	\dot n(\Omega_j) \equiv 0 \,\,\, {\rm if}\,\,\,  n(\Omega)=n_B(\Omega)\equiv \frac{1}{e^{\beta\hbar\Omega}-1}.
\end{equation}
This can be seen, since the following identities hold for $n_B(\Omega)$: $n_B(\Omega_j+\Omega_k) = n_B(\Omega_j)n_B(\Omega_k)/[1+n_B(\Omega_j)+n_B(\Omega_k)]$, and $n_B(\Omega_j-\Omega_k) = n_B(\Omega_j)[1+n_B(\Omega_k)]/[n_B(\Omega_k)-n_B(\Omega_j)]$. Based on these relations, all terms in the sums on the right-hand side of Eq. \eqref{kinfull} vanish identically for any temperature $T=(k_B \beta)^{-1}$. Below we will find the actual value of $T$ based on energy conservation. 

We can test whether the solution $n_B(\Omega)$ is stable by introducing small deviations $\delta n(\Omega_j)$ around this stationary solution. We then find that
\begin{widetext} 
\begin{equation}\label{kinstab}
	\frac{d\delta n(\Omega_j)}{dt} = -\frac{4\pi}{\hbar^2}\nu_0\big\{\sum_{k=1}^N 2|A_{jk}|^2|B_{j+k}|^2\frac{n_B(\Omega_k)[1+n_B(\Omega_k)]}{1+n_B(\Omega_j)+n_B(\Omega_k)}	 +\sum_{k=1}^{j-1} |A_{k,j-k}|^2|B_j|^2 \frac{n_B(\Omega_k)[1+n_B(\Omega_k)]}{n_B(\Omega_k)-n_B(\Omega_j)} \big\}\delta n(\Omega_j).
\end{equation}
\end{widetext}
This equation can then be written as $\frac{d\delta n(\Omega_j)}{dt} = -\kappa_{j}\delta n(\Omega_j)$, where $\kappa_j$ is positive for any non-vanishing coupling of form \eqref{threewave}, securing stability of the stationary distribution. Numerical examples of the dynamics of $n(\Omega_j)$ will be given below, further demonstrating the stability of the distribution.

The remaining question is how the temperature of the system is determined. This in fact can be answered based on an elementary energy conservation argument for an isolated system such as ours. The internal energy of the initial state is
\begin{equation}\label{Uini}
	U_{\rm ini}=\sum_{j=1}^N N_{j}^{(\rm ini)}\hbar \Omega_j,
\end{equation}
where $N_{j}^{(\rm ini)}$ represents the initial population in state $j$. If one eventually at $t \rightarrow \infty$ reaches the state with populations $n_B(\Omega_j)$, the final internal energy is
\begin{equation}\label{Ufin}
	U_{\rm fin}=\sum_{j=1}^N  \frac{\hbar \Omega_j}{e^{\beta \hbar\Omega_j}-1}.
\end{equation}
Equating $U_{\rm fin}=U_{\rm ini}$, we then find $\beta$. For a sufficiently large initial energy the final state spreads over many modes, and we can approximate the sum in Eq. \eqref{Ufin} by an integral, with the result
\begin{equation}\label{Ufina}
	U_{\rm fin}\approx \frac{\pi^2}{6\hbar \Delta\Omega\beta^2}.
\end{equation}
Then, for instance, if we initialize the system with an excitation on the level $k_{\rm ini}$, the temperature is given approximately by
\begin{equation}\label{T}
\beta\hbar\Delta \Omega \approx \pi/\sqrt{6 k_{\rm ini}}.
\end{equation}

 \begin{figure}
	\centering
	\includegraphics [width=0.5\textwidth] {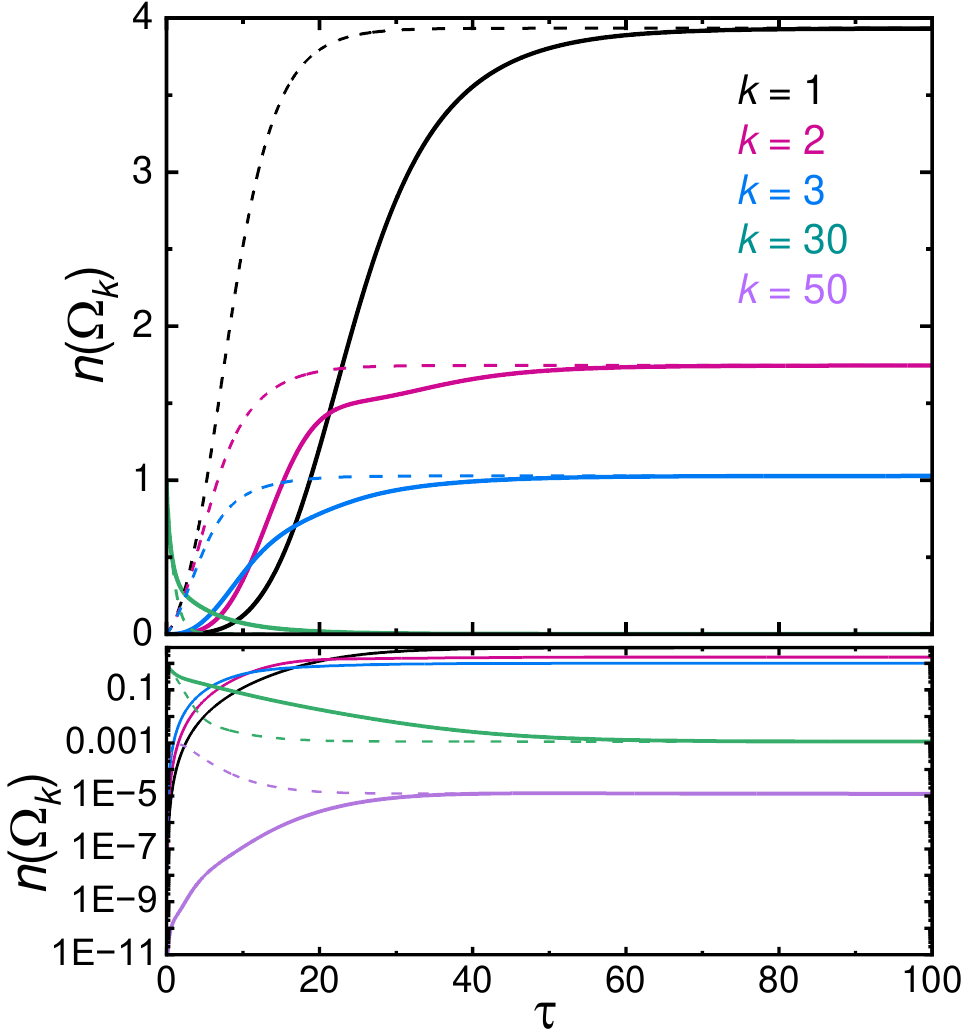}
	\caption{Time-dependent populations on various photon modes $k$ of the multimode resonator when initialized on the $k_{\rm ini}=30$ level with $N_{30}=1$ at $\tau\equiv \Delta \Omega t=0$. The solid lines correspond to the coupling of form $|A_{pq}|^2=g^2 e^{-\gamma|p-q|},\, |B_r|^2=r$ \cite{manucharyan2023}, and $\Gamma_0\equiv \frac{4\pi}{\hbar^2}\nu_0 g^2 =0.01\Delta\Omega$, with $\gamma=1$. The dashed lines on the other hand correspond to the coupling  $\frac{4\pi}{\hbar^2}|A_{pq}|^2|B_r|^2 = 0.03$ for all $p,q,r$.  
		\label{fig2}}
\end{figure}
Next we present numerical results of the dynamics of the system. First we analyze the time dependence of the populations in different modes after the system is intialized in a given state. Figure \ref{fig2} is an example of such dynamics, presenting the case of system initially occupying the mode $k_{\rm ini} = 30$. Populations at different time instants $\tau=\Delta \Omega t$ of modes $k=1,2,3,30$ and $50$ are shown. We assume that there are $N=800$ lowest modes available. In the calculation we assume that the three-wave coupling is parametrized as $|A_{pq}|^2=g^2 e^{-\gamma|p-q|},\, |B_r|^2=r$ \cite{manucharyan2023}, and $\Gamma_0\equiv \frac{4\pi}{\hbar^2}\nu_0 g^2 =0.01\Delta\Omega$, with $\gamma=1$. We see that the system finds its asymptotic state at times $\tau \gg 100$ in this case. Very similar dynamics can be found with other choices of the coupling parameters, for instance assuming constant $|A_{pq}|^2$ and $|B_r|^2$ as demonstrated by the second set of curves (dashed lines) in Fig. \ref{fig2}. To highlight the rarely populated states, we plot $n(\Omega_k)$ for the same states on a logarithmic scale in the lower panel of Fig. \ref{fig2}.

 \begin{figure}
	\centering
	\includegraphics [width=0.5\textwidth] {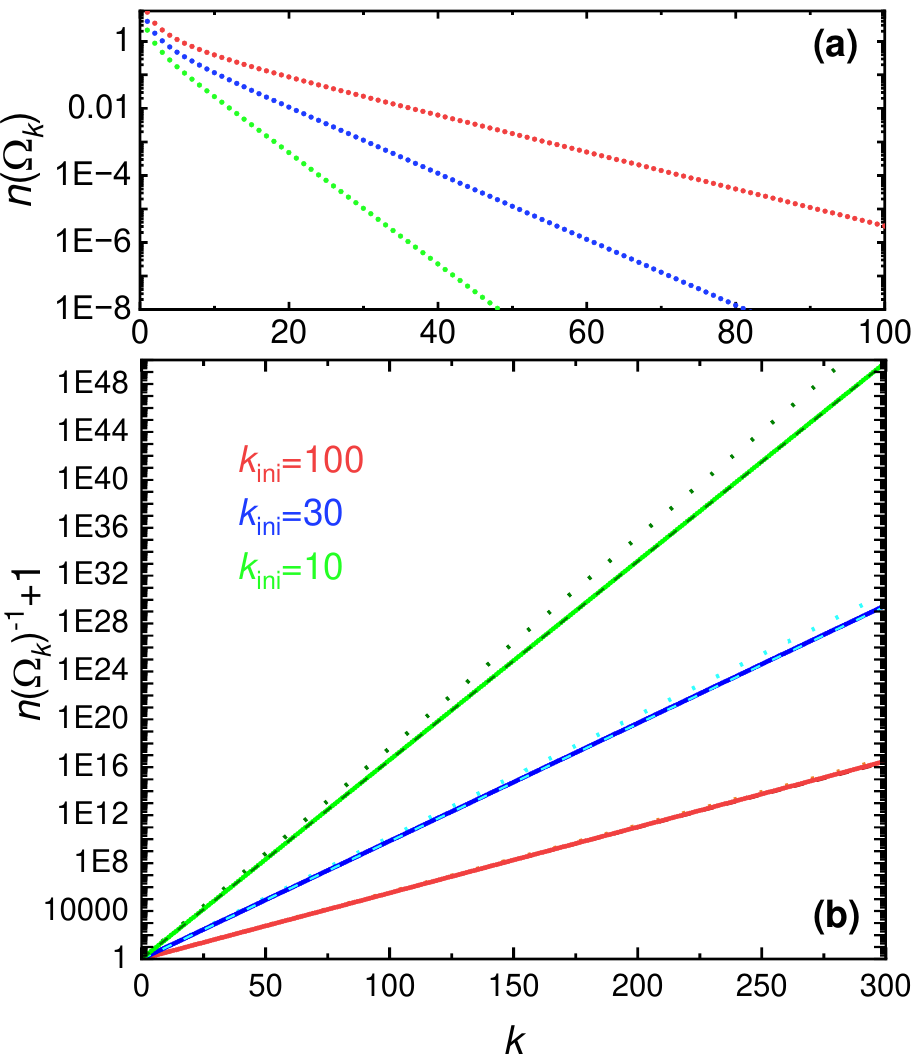}
	\caption{Long term asymptotic distribution of oscillator mode populations in form $n(\Omega_k)^{-1}+1$ for three different initial states, $k_{\rm ini}=10,30$ and $100$. The symbols are from the numerics, the dashed lines are the exact Bose-Einstein distributions assuming energy conservation, and the dotted line is the approximate result based on Eq. \eqref{T}.
		\label{fig3}}
\end{figure}
Based on the numerical evolution in Fig. \ref{fig2} we expect the system to be stable and find a state which is potentially the given steady state solution $n_B(\Omega)$. We may test the analytical predictions by plotting the distribution at a time where populations are stable in time, here in Fig. \ref{fig3} at $\tau=150$ for the couplings with the same set of parameters as above. In Fig. \ref{fig3} (a) we present these asymptotic populations $n(\Omega_k)$ vs $k$. To test the results against the predicted ones, we plot in Fig. \ref{fig3} (b) $n(\Omega_k)^{-1}+1$ vs $k$ for three different initial values of $k_{\rm ini}=10,30$ and $100$, demonstrating exponential dependence over nearly 50 orders of magnitude. For comparison, we show $n_B(\Omega_k)^{-1}+1=e^{\beta \hbar\Omega_k}$ where $\beta$ is obtained by equating $U_{\rm ini}$ from Eq. \eqref{Uini} with $U_{\rm fin}$ either from Eq. \eqref{Ufin} (accurate, dashed line) or from Eq. \eqref{T} (integral approximation, dotted line). The first one coincides perfectly with the numerical $n(\Omega_k)^{-1}+1$ over all $k$, whereas the approximation becomes less accurate for lower $k_{\rm ini}$ as expected.

Discussion: Our non-linear system loses information of its initial state, apart from its total energy, thus exhibiting full thermalization. This is contrary to several other model systems, like a one-dimensional chain of hard-core bosons \cite{rigol2007} or quadratically coupled two-level systems \cite{pekola2023}. They relax to their maximum entropy state \cite{rigol2007,ilari2024} according to the principle of Generalized Gibbs Ensemble \cite{jaynes1957,rigol2007,ilari2024}, still remembering the initial state even at long times due to, e.g., the excitation number (and energy) conservation in the latter model. 
%A quadratically coupled system is integrable due to excitation number conservation, but it relaxes to its maximum entropy state \cite{ilari2024,pekola2023} according to the principle of Generalized Gibbs Ensemble \cite{jaynes1957,rigol2007}. Having a lot in common with our present model, it is expected that a three-wave coupling would introduce similar further relaxation towards a true thermal distribution at later times, thus exhibiting the full prethermalization-thermalization sequence as discussed widely on general arguments \cite{mori2018}. Our weak-coupling model (Eq. \eqref{kin1}) does not capture the prethermalization step but retains the initial population instead of reaching a Lorentzian distribution around the initial state \cite{pekola2023}. 
In our analysis we employed the perturbative approach (Kubo formula, Eq. \eqref{kin1}) and the continuum approximation of states when securing the energy conservation in the step between Eqs. (12) and (13) in SM. The implication of the latter approximation is still not quite clear to us, and it remains to be seen in further work, whether the true thermalization without Poincare recoveries can be achieved only in multimode resonators with very small free spectral range $\Delta \Omega$ \cite{zhong2019, qiu2023}, and what determines this condition. We find an analogous situation in the problem of electron-electron relaxation \cite{pothier1997}, where in metallic samples the density of states is large, of the order of inverse 1 mK, due to the large number of electrons, securing internal thermalization. Similarly, a long (1 - 100 m) superconducting waveguide yields a free spectral range of the same order. Moreover, superconducting waveguides can have quality factors in excess of $10^5$, lending a regime of quasi-isolation over times of order 1 s. Therefore long nonlinear superconducting waveguides \cite{qiu2023} could serve as a testbed for our model. In summary, solid state systems, in particular superconducting quantum circuits provide a new, highly controllable pathway to the studies of thermalization in quantum regime.

Acknowledgements: We thank Joachim Ankerhold, Andrew Cleland, Ilari M\"akinen, Teemu Ojanen and Ciprian Padurariu for useful discussions. This work
was financially supported by the Research Council of Finland Centre of Excellence program grant 336810 and grant
349601 (THEPOW).

\cleardoublepage % just to show that the cross-references go to different places

\appendix

\counterwithin*{figure}{part}

\stepcounter{part}

\renewcommand{\thefigure}{A.\arabic{figure}}

		\section{Supplemental Material}

\section{Relaxation of a multilevel system by three-wave mixing}
The total Hamiltonian of a multilevel system (cavity) reads
\begin{equation}\label{Hamilton1}
	\hat{\mathcal{H}}=\hat{\mathcal{H}}_{0}+\hat V,
\end{equation}
where $\hat{\mathcal{H}}_{0}=\sum_{j=1}^N \hbar \Omega_j a_j^\dagger a_j$ is the non-interacting Hamiltonian. 
\begin{equation} \label{coupling}
	\hat V=\sum_{k,l,m} (M_{klm}a_k^\dagger a_l^\dagger a_m +M_{klm}^* a_m^\dagger a_l a_k)
\end{equation} 
presents the three-wave coupling Hamiltonian between the cavity modes. Number operator for the $j$:th level of the cavity is given by $\hat{N}_j=a_j^\dagger a_j$ and its expectation value is denoted by $n(\Omega_j)\equiv \langle \hat{N}_j\rangle$. In order to find the kinetic equation for $n(\Omega_j)$, we start with the time rate of change of the operator $\hat{N}_j$ given by
\begin{widetext}
\begin{equation}\label{Ndotj-1}
	\dot {\hat N}_j =\frac{\imath}{\hbar}[\hat V,\hat N_j]=\frac{\imath}{\hbar}\sum_{k,l,m} \big{\{}M_{klm}[a_k^\dagger a_l^\dagger a_m ,a_j^\dagger a_j]+M_{klm}^*[a_m^\dagger a_l a_k,a_j^\dagger a_j]\big{\}}.
\end{equation}
\end{widetext}
For bosonic modes where we have $[a_i,a_j^\dagger]=\delta_{ij}$, Eq.~\eqref{Ndotj-1} reads
\begin{widetext}
\begin{equation}\label{Ndotj-2}
	\dot {\hat N}_j =\frac{\imath}{\hbar}\sum_{k,l} \big{\{}M_{klj} a_k^\dagger a_l^\dagger a_j-M_{kjl}a_j^\dagger a_k^\dagger a_l-M_{jlk}a_j^\dagger a_l^\dagger a_k+M_{jlk}^*a_k^\dagger a_l a_j+M_{kjl}^*a_l^\dagger a_k a_j-M_{klj}^*a_j^\dagger a_l a_k\big{\}}.
\end{equation} 
\end{widetext}
Next we use the Kubo formula in the interaction picture as
\begin{equation}\label{kin-1}
	\dot n(\Omega_j) =-\frac{\imath}{\hbar}\int_{-\infty}^0 dt'\langle [\dot {\hat N}_j,\hat V_I(t')]\rangle_0,
\end{equation}
where $\hat V_I(t) = e^{i\hat{\mathcal{H}}_{0}t/\hbar} \hat V e^{-i\hat{\mathcal{H}}_{0}t/\hbar}$, and $\langle \cdot\rangle_0$ denotes averaging in the non-interacting system. Substituting the expresions of Eq.~\eqref{Ndotj-2} and the coupling term in the interaction picture in Eq.~\eqref{kin-1}, we will have twelve terms. Here, we use one of these terms as an example to illustrate the derivation of the complete expression for $\dot n(\Omega_j)$. This "first" term in the integrand of Eq. \eqref{kin-1} reads
\begin{widetext}
\begin{eqnarray}\label{1st-term-1}
	1^{\rm st}=\frac{1}{\hbar^2}\sum_{k,l}\sum_{k',l',m'}M_{klj}M_{k'l'm'}^*\langle a_k^\dagger a_l^\dagger a_j a_{m'}^\dagger a_{l'} a_{k'}\rangle e^{\imath(\Omega_{m'}-\Omega_{l'}-\Omega_{k'})t'}.
\end{eqnarray}
\end{widetext}
Applying the Wick's theorem for the expectation values of bosonic operators we have
\begin{widetext}
\begin{eqnarray}\label{1st-term-2}
	1^{\rm st}=&&\frac{1}{\hbar^2}\sum_{k,l}\sum_{k',l',m'}M_{klj}M_{k'l'm'}^*\big{\{}\langle a_k^\dagger a_j\rangle \langle a_l^\dagger a_{l'}\rangle\langle a_{m'}^\dagger a_{k'}\rangle + \langle a_k^\dagger a_j\rangle \langle a_l^\dagger a_{k'}\rangle\langle a_{m'}^\dagger a_{l'}\rangle\\&& +\langle a_k^\dagger a_{l'}\rangle \langle a_l^\dagger a_{j}\rangle\langle a_{m'}^\dagger a_{k'}\rangle + \langle a_k^\dagger a_{l'}\rangle \langle a_l^\dagger a_{k'}\rangle\langle  a_{j} a_{m'}^\dagger \rangle + \langle a_k^\dagger a_{k'}\rangle \langle a_l^\dagger a_{j}\rangle\langle a_{m'}^\dagger a_{l'}\rangle + \langle a_k^\dagger a_{k'}\rangle \langle a_l^\dagger a_{l'}\rangle\langle  a_{j} a_{m'} \rangle\big{\}}e^{\imath(\Omega_{m'}-\Omega_{l'}-\Omega_{k'})t'}.\nonumber
\end{eqnarray}
\end{widetext}
Since we have $\langle a^\dagger a\rangle=n(\Omega)$ and $\langle a a^\dagger\rangle=[1+n(\Omega)]$, Eq.~\eqref{1st-term-2} reads
\begin{widetext}
\begin{eqnarray}\label{1st-term-3}
	1^{\rm st}=&&\frac{1}{\hbar^2}\big{\{}\sum_{l,k'}M_{jlj}M_{k'lk'}^*\,\, n(\Omega_j)n(\Omega_l)n(\Omega_{k'})e^{-\imath \Omega_l t'}+ \sum_{l,l'}M_{jlj}M_{ll'l'}^*\,\, n(\Omega_j)n(\Omega_l)n(\Omega_{l'})e^{-\imath \Omega_l t'}\nonumber\\&&~~~ +\sum_{k,k'}M_{kjj}M_{k'kk'}^*\,\, n(\Omega_k)n(\Omega_j)n(\Omega_{k'})e^{-\imath \Omega_k t'}+ \sum_{kl}M_{klj}M_{lkj}^*\,\, n(\Omega_k)n(\Omega_l)[1+n(\Omega_{j})]e^{\imath (\Omega_j-\Omega_k-\Omega_l) t'}+ \nonumber\\&&~~~\sum_{kl'}M_{kjj}M_{kl'l'}^*\,\, n(\Omega_k)n(\Omega_j)n(\Omega_{l'})e^{-\imath \Omega_k t'}+\sum_{kl}M_{klj}M_{klj}^*\,\, n(\Omega_k)n(\Omega_l)[1+n(\Omega_{j})]	\big{\}}e^{\imath(\Omega_{j}-\Omega_{l}-\Omega_{k})t'}.
\end{eqnarray}
\end{widetext}
By following the same procedure, we can derive the remaining eleven terms. Combining them all and substituting into Eq.~\eqref{kin-1}, and carrying out the integral, we obtain 
\begin{widetext}
\begin{eqnarray}\label{kin-2}
	\dot n(\Omega_j) =\frac{4\pi}{\hbar^2}\sum_{k,l}&& \big{\{}2|A_{jk}|^2|B_l|^2\,\big{[} n(\Omega_l)+n(\Omega_j)n(\Omega_l)+n(\Omega_k)n(\Omega_l)-n(\Omega_j)n(\Omega_k)\big{]}\delta (\Omega_l-\Omega_j-\Omega_k)\\
	&&-|A_{kl}|^2|B_j|^2\,\big{[}n(\Omega_j)+n(\Omega_j)n(\Omega_k)+n(\Omega_j)n(\Omega_l)-n(\Omega_k)n(\Omega_l)\big{]}\delta (\Omega_j-\Omega_k-\Omega_l)\big{\}},\nonumber
\end{eqnarray}
\end{widetext}
where we express the coupling term in the form $M_{jkl} \equiv A_{jk}B_l$. Replacing $\sum_{l}$ on all possible modes by the integral $\sum_{l}\rightarrow \nu_0\int_0^{\Omega_{\rm max}} d\Omega_{l}$, where $\nu_0$ is the density of modes and $\Omega_{\rm max}$ is the frequency of the highest mode considered, we have finally
\begin{widetext}
\begin{eqnarray}\label{kinfull}
	\dot n(\Omega_j) = \frac{4\pi}{\hbar^2}\nu_0\big{\{}&&\sum_{k=1}^N 2|A_{jk}|^2|B_{j+k}|^2\,\big{[}n(\Omega_j+\Omega_k)+n(\Omega_j)n(\Omega_j+\Omega_k)+n(\Omega_k)n(\Omega_j+\Omega_k)-n(\Omega_j)n(\Omega_k)\big{]}	\nonumber\\&&-\sum_{k=1}^{j-1} |A_{k,j-k}|^2|B_j|^2\,\big{[}n(\Omega_j)+n(\Omega_j)n(\Omega_k)+n(\Omega_j)n(\Omega_j-\Omega_k)-n(\Omega_k)n(\Omega_j-\Omega_k)\big{]}\big{\}}.
\end{eqnarray}
\end{widetext}
\section{Quadratic coupling}
If we assume quadratic coupling 
\begin{equation}\label{coupling2}
	\hat V=\frac{1}{2}\sum_{p\neq q} g_{pq}(a_p+a_p^\dagger)(a_q+a_q^\dagger),
\end{equation} 
instead of the three-wave mixing (Eq. \eqref{coupling}), we find by a similar analysis
\begin{equation}\label{Ndotj-2nd}
	\dot {\hat N}_j =\frac{\imath}{\hbar}\sum_{p} (g_{pj}+g_{jp})(a_pa_j+a_p^\dagger a_j-a_j^\dagger a_p-a_j^\dagger a_p^\dagger).
\end{equation} 
With Eq. \eqref{kin-1} we then obtain
\begin{equation}\label{norel}
	\dot n(\Omega_j) =\frac{2 \pi}{\hbar^2}\sum_{p} (g_{pj}+g_{jp})^2[n(\Omega_p)-n(\Omega_j)]\delta(\Omega_p-\Omega_j) =0,
\end{equation} 
demonstrating absence of relaxation in this case.

\end{document}